\documentstyle[12pt]{article}
\title{How Many New Worlds Are Inside a Black Hole?}
\date{}
\author{Claude Barrab\`{e}s\thanks{Electronic address:
barrabes@univ\_tours.fr}\\
Laboratoire de Mod\`{e}les de Physique Math\'{e}matique\\
Universit\'{e} de Tours, 37200 Tours, France\\
and D\'{e}partement d'Astrophysique Relativiste et Cosmologie\\
Observatoire de Paris, 92190 Meudon, France\bigskip\\
Valeri P. Frolov\thanks{Electronic address:
frolov@phys.ualberta.ca}\\
CIAR Cosmology Program, Theoretical Physics Institute\\
University of Alberta, Edmonton, Canada T6G 2J1\\
and P.N.Lebedev Physics Institute, Moscow, Russia}

\begin{document}
\maketitle

\begin{abstract}
We propose a possible internal structure for a Schwarzschild black
hole
resulting from the
creation of multiple de Sitter universes with lightlike boundaries
when the
curvature
reaches Planckian values. The intersection of the boundaries is
studied and a
scenario
leading to disconnected de Sitter universes is proposed. The
application to the
information
loss problem is then discussed.
\end{abstract}

\centerline{PACS numbers: 04.60.+n, 03.70.+k, 98.80.Hw}
\newpage
\setcounter{page}{1}

\section{Introduction}
The internal structure of black holes and their final state are
two
intriguing
problems of black
hole physics. Both of these problems require knowledge of physics
at
Planckian
scales for their
solution. It is generally believed that only a union of quantum
mechanics
and gravity
can provide us with a proper theory. Till now such a theory,
quantum gravity,
has not been
constructed. It seems that one cannot overcome its main problem
---
non-renormalizability --- without unifying gravity with other
physical
fields. Superstring
theory is one of the most promising approaches in this direction.
But
in spite
of the
impressive development of  superstring theory we are still very
far from
understanding physics at Planckian scales.

Under these circumstances it is natural to use the following
approach. One
might assume
that the notion of quantum average of a metric $g=\langle
\hat{g}\rangle$ is
still valid in the
regions under consideration, and the average metric $g$ obeys some
effective
equations.
We do not know these equations at the moment, but we might assume
that these
equations
and their solutions obey some general properties and restrictions.
For example,
it is natural
to require that the effective equations for $g$ in the low
curvature
limit
reduce to the
Einstein equations with possible higher-curvature corrections. It
is also
possible to
assume that the future theory of quantum gravity would  solve the
problem of
singularities
of classical General Relativity.
One of the possibilities is that the equations of the complete
theory would simply not allow dynamically infinite growth of the
curvature, so
that the
effective curvature  $\cal R$ of $g$ is bounded by the Planckian
value
$\approx
1/l_{Pl}^2$. The principle of a  limiting curvature was
proposed by Markov \cite{Mark:82,Mark:84}. This principle excludes
curvature
singularity formation, so that  the global properties of the
solutions  must
change.

A special form of the gravitational action for cosmological models
providing
the limiting curvature principle  was considered in
Refs.\cite{MaMu:85}. It was
shown that a collapsing homogeneous isotropic universe must stop
its
contraction and begin expansion, while during the transition phase
its
evolution is described by a metric close to the de Sitter one.
Brandenberger
and Mukhanov \cite{MuBr:92} proposed a general non-linear
gravitational  action
which allows only regular homogeneous isotropic solutions.
Polchinski
\cite{Polc:89} proposed a simple realization of the limiting
curvature
principle by modifying the action and inserting inequality
constraints into it,
 restricting the  growth of  curvature.  In the case of the
collapse of an
inhomogeneous universe, formation of a few 'daughter universes' can
be
expected \cite{Mark:84}.

 In the application to the black-hole-interior problem the
limiting-curvature
principle means that the singularity which, according to the
classical theory exists  inside a black hole, must be removed in
the
complete
quantum theory, so that the global structure of spacetime would be
essentially modified.
We cannot hope to derive this result without  knowledge of the
theory, but
we may at least
discuss and classify  possibilities. Such a 'zoological' approach
is
a natural
first step and it
was used in a number of publications.
One of the first models of a spherically-symmetric black hole
without
singularities was
proposed by Frolov and Vilkovisky \cite{FrVi:81}. In this model the
apparent
horizon does not cross $r=0$, so that $r=0$ is a regular time-like
line. Two
cases are
logically possible. (1) Inner and  external parts of the apparent
horizon remain disparate.
In this case a black hole does not evaporate completely and
a permanent
black-hole remnant remains. (2) The apparent horizon is closed. In
this case,
there is no
event horizon (and hence,  strictly speaking no black hole), but
practically
all the
observable properties of a black hole would be present until  late
times,
when the
apparent horizon disappears. This model was discussed later in
Refs.\cite{RoBe:83} and, recently, both types of these
 singularity-free models of a black hole were used in the
discussion
of the
final state of an evaporating black hole and information loss
problem
(see e.g.
\cite{Stro:92}).

Another logically possible singularity-free model of a black-hole
interior
was proposed by  Frolov, Markov, and Mukhanov
\cite{FrMaMu:89,FrMaMu:90}.
According to this model, inside a black hole there  exists a closed
Universe
instead of a singularity .
The metric is obtained by gluing the Schwarzschild metric  to the
de
Sitter
metric through a surface $r=r_0$=const  located
inside the black hole. The parameter $r_0$ is chosen in such a way
that  the
value of  a curvature calculated at $r_0$ coincides with a limiting
curvature,
which is assumed to be of  Planckian order. In this
approach a
fast transition between regimes is assumed  and the transition
region
required
for change of the regimes is  approximated by a thin spacelike
shell. In the FMM-model
\cite{ftn1}  the spacetime passes through the deflation stage and
instead of
the singularity a
new inflating Universe is created.
Morgan\cite{Morg:91} showed that a similar result (formation of a
contracting
closed de Sitter-like universe with its further inflation), can be
obtained in
the framework of the Polchinski approach to the limiting curvature
principle.
Different aspects of the model of a singularity-free black hole
interior with
an inner de Sitter
like universe  were also studied
in\cite{Polc:89,Morg:91,IsPo:88,BaPo:90}.

One of the  assumptions of FMM and other similar models is that a
'phase
transition' to the de Sitter-like phase takes place  at the
homogeneous
spacelike surface $r=r_0$.  The presence of perturbations and
quantum
fluctuations, growing as $r\rightarrow 0$, could  spoil the
homogeneity. The
bubbles of the new de Sitter-like phase could be formed
independently
at points
separated by spacelike distances. For these reasons one could
expect
that
different parts of a black-hole interior can create spatially
disconnected
worlds. The aim of this paper is to consider a simple model which
could
describe possible  features of this process. Namely we suppose that
spherical
bubbles of the new de Sitter-like phase which are formed
independently are
separated
from the old (Schwarzschild) phase  by relativistically moving
boundaries.
Under this assumption one can  reduce the problem to the study of
the evolution of
light shells representing the boundaries,
and their intersection.  The general theory of lightlike
shells was
developed by Barrabes and Israel\cite{BaIs:91} (see
also\cite{BeKuTk:90}).
This approach is purely kinematic in the following sense. It allows
one to
take into
account the conservation of energy and momenta during the process
of
nucleation and the
further evolution of the boundaries, including possible
intersection
of the
boundaries of two different bubbles. But it certainly does not
answer
questions concerning probability of bubble formation or the
structure of the
transition regions between phases. One of the interesting results
of
the model
is that it does not exclude creation of a large  number of new-born
universes.
This fact could have an interesting application to the
information-loss puzzle.

The paper is organized as follows. Section 2 contains the
discussion
of the
model and gives the
conditions for the nucleation of a de Sitter bubble inside a
Schwarzschild
black
hole. The creation of multiple
de Sitter bubbles is considered in Section 3, and the
interaction
between the boundaries of  newly created de Sitter bubbles is
discussed in
Section 4. In Section 5 we discuss the possible application of the
process of
multiple universe  formation to the information loss puzzle.
Finally some of the main properties of timelike and
lightlike shells which are used in our model are summarized
in Appendix.

\section{Description of the Model}
\subsection{A model}

The Schwarzschild metric
\begin{equation} \label{eq1.1}
ds^2=-F^{-1}dr^2+Fdt^2 +r^2 d\Omega^2,
\end{equation}
\begin{equation}
F=r_+/r -1 ,\hspace{1cm}d\Omega^2=d\theta^2 + \sin^2\theta d\phi^2
\end{equation}
inside the gravitational radius ($r<r_+$) describes the contracting
homogeneous
Kasner-like universe with the isometry group
$R(1)\times O(3)$. The section of fixed time $r=\mbox{const}<r_+$
is
a
homogeneous
spacelike surface $\Sigma$ with topology $R^1\times S^2$. The
square
of the
curvature ${\cal R}^2\equiv
R_{\alpha\beta\gamma\delta}R^{\alpha\beta\gamma\delta}=12
r_+^2/r^6$.
The value
of the curvature $\cal R$ is
constant along $\Sigma$ and it is of order of the Planckian
curvature
$l_{Pl}^{-2}$
at $r\sim r_0 = (12)^{1/6} (r_+/l_{Pl})^{1/3}l_{Pl}$. In the
FMM-model it is
assumed that as soon as the spacetime curvature reaches some
critical
value
${\cal R}=1/l^2$ a new de Sitter-like phase is formed. In the
application to
the unperturbed Schwarzschild metric the change to a new phase
occurs everywhere
simultaneously (at the spacelike surface $r=r_0\sim (r_+
l^2)^{1/3}$). Another
assumption of the FMM-model is that the transition takes a short
time,
so that
the transition region can be approximated by a thin shell.

For a black hole formed by the collapse of a body with small
deviations from
spherical symmetry  the metric at a finite
radius $r$  tends to the
Schwarzschild metric (\ref{eq1.1}) at large  distance from the
collapsing
body\cite{DoNo:78}.
On the other hand   perturbations existing in the
black hole exterior and propagating inwards
grow infinitely
near the singularity.
Quantum fluctuations of metric also become important when the
spacetime curvature reaches the limiting (Planckian) value
$l^{-2}$.
The
'phase transition' into the new de Sitter-like phase happens
independently in
different spatially separated  parts of the black hole interior.
It is
also
plausible that due to the fluctuations of the new-phase bubble
formation there
is a  dispersion in the  times  of bubble formation.   Under
these
conditions the assumption of spatial homogeneity used in the
FMM-model is
rather
restrictive and it is necessary to consider a  generalization of
this model.
Our purpose is to generalize the FMM model to the case where the
transition to
the de
Sitter-like phase occurs independently in spatially separated
regions. In order
to describe the possible structure of spacetime we consider a toy
model in
which spherical symmetry is preserved. We assume that, near the
singularity of
a Schwarzchild black hole, the transition to a new 'de Sitter
phase'
takes
place at  two-spheres
$S$. We preserve another assumption of the FMM-model, namely that
the
transition takes a short time and the transition region can be
approximated by a
thin shell. In our generalization of the FMM-model we assume that
the
boundary
between the two phases (Schwarzschild and de Sitter) is composed of
two null
hypersurfaces lying to the future of $S$ and intersecting at $S$.
Using this simple model we
discuss different possibilities of nucleation of bubbles with de
Sitter-like
interiors and the interaction between the newly created bubbles.

\subsection{Lightlike shells}
Lightlike shells separating two regions of a spacetime with
different
characteristics have proved to be a convenient way of
dealing with various physical or  mathematical problems in general
relativity
\cite{BaIs:91}.
This happens because the dynamics of lightlike shells is simple and
it
is directly
related with geometrical properties at the junction of the two
spacetime
regions. A good
example
illustrating this  is the collision of two null shells. It has been
shown,
first in the restricted case of spherical symmetry
\cite{Redm:85,ClDr:87}
and later in
more general
situations \cite{BaIsPo:90} that the geometries of the four
spacetime
domains
bounded by the
ingoing and outgoing shells are matched at the collision by only
two
remarkably
simple
algebraic relations. In the particular case of two concentric
spherical shells
moving radially
toward each other with the velocity of light one of the matching
conditions is
trivially satisfied while the other takes the form
\begin{equation} \label{eq2.1}
f_A(r_0) f_B(r_0) = f_C(r_0) f_D(r_0).
\end{equation}
Hear $r_0$ is the radius of the collision sphere, and the functions
$f_A, f_B,
\ldots$ for the spacetime domains $A, B, \ldots$ [see Fig
1.a] are defined by $f = g^{\alpha \beta} \partial_{\alpha}r
\partial_{\beta}r
= g^{rr}$. In what follows we consider a spherically symmetric
spacetime with
the metric
\begin{equation} \label{eq2.2}
ds^2 = - f(r) dt^2 + f^{-1}(r) dr^2 + r^2 d\Omega^2.
\end{equation}
Both Schwarzschild and de Sitter metrics are of this
form\footnote{In
 general
a spherically symmetric solution of Einstein equation is of the
form
(\ref{eq2.2}) provided the stress-energy tensor obeys the condition
${T^r}_r =
{T^t}_t$.}.

\begin{figure}
\let\picnaturalsize=N
\def\picsize{10.0cm}
\def\picfilename{fig1.eps}
\ifx\nopictures Y\else{\ifx\epsfloaded Y\else\input epsf \fi
\let\epsfloaded=Y
\centerline{\ifx\picnaturalsize N\epsfxsize \picsize\fi
\epsfbox{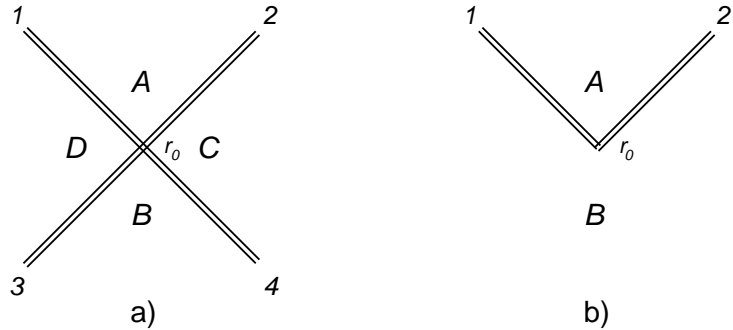}}}\fi
\caption{The intersection of two lightlike shells (Fig.1.a). The
incoming
lightlike shells 3 and 4 after intersection at the two-dimensional
surface
$r_0$ propagate further as lightlike shells 1 and 2 with
parameters,
different from the original ones. A special case of this process is
shown in
Fig.1.b : the masses of the in-coming shells 3 and 4 vanish, so
that
out-going
shells 1 and 2 are created from 'nothing'.}
\end{figure}

As a special case the formula (\ref{eq2.1}) describes the creation
of
a pair of
lightlike shells
from 'nothing' [see Fig 1.b]. When two lightlike shells
are created at the sphere of  radius $r_0$ the geometries of the
three
spacetime
domains $B, C, D$ are identical and   the Eq. (\ref{eq2.1}) becomes
\[
f_B(r_0) \left( f_A(r_0) - f_B(r_0) \right) = 0.
\]
If none  the  shells coincides with the horizon of the region $B$ (
$f_B(r_0)
\ne 0$), the matching relation (\ref{eq2.1}) reduces to\footnote{It
should be
stressed that this relation is  valid only at the surface of
collision. Similar
relations can be easily obtained for the time reversed process
(annihilation of
two lightlike shells) and for the
bounce of a lightlike shell --- see examples in
Ref.\cite{BaIs:91}.}
\begin{equation} \label{eq2.3}
f_A(r_0) = f_B(r_0).
\end{equation}
The consistency between the geometrical formula (\ref{eq2.3}) and
the
conservation laws
which have to be satisfied at the moment of creation of two shells
can be
checked.

In a spherically symmetric spacetime and in the absence of energy
fluxes the
surface stress-energy tensor of the null shell is uniquely
determined
by the
surface energy  density $\sigma(r)$, which is  given by
\begin{equation} \label{eq2.4}
4 \pi r^2 \sigma(r) = \zeta r \left[f_{+}(r) - f_{-}(r)\right]/2 .
\end{equation}
Here $\zeta = +1 (-1)$ if $r$ increases (decreases) in the
direction
of the
future-directed
null generators $n = \zeta \partial / \partial r$ and  $f_{+}$ $
(f_{-})$
refers to the future
(past) side of the shell (for more details see  Ref.\cite{BaIs:91}
and
Appendix A ).
The relations  (\ref{eq2.3}) and (\ref{eq2.4}) show that the energy
surface
density of the lightlike shells must vanish at the moment of
creation
($\sigma_1(r_0) = \sigma_2(r_0) = 0$). After the creation the
shells
possess
non-vanishing  surface energy densities $\sigma_1(r)$ and
$\sigma_2(r)$ given
by (\ref{eq2.4}). The relative signs of $\sigma_1(r)$ and
$\sigma_2(r)$ depend
on the   values of the $\zeta$'s and the jump of the function $f$.

\subsection{De Sitter-phase bubble creation in the  black hole
interior}
We consider at first the formation of a single de Sitter-phase
bubble.
Denote by $r_0$
the radius of the sphere $S$ where  a single bubble is nucleated.
We
call $S$ the
{\em vertex sphere}.  According to our assumption the creation of
the
bubble is
accompanied by the creation of two  lightlike shells separating a
newly formed
de Sitter-phase from the 'old' Schwarzschild one.
The corresponding Penrose-Carter diagram is shown in Fig.2.

\begin{figure}
\let\picnaturalsize=N
\def\picsize{10cm}
\def\picfilename{fig2.eps}
\ifx\nopictures Y\else{\ifx\epsfloaded Y\else\input epsf \fi
\let\epsfloaded=Y
\centerline{\ifx\picnaturalsize N\epsfxsize \picsize\fi
\epsfbox{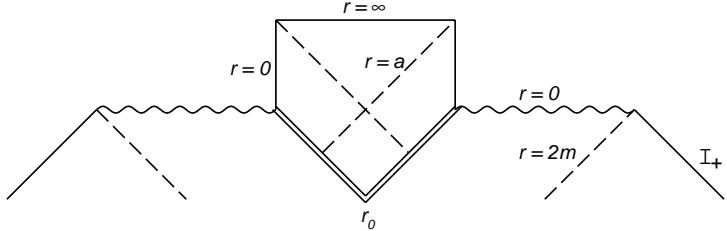}}}\fi
\caption{A single de Sitter-phase bubble formation.}
\end{figure}

Both shells   converge towards $r=0$. For this reason  their
creation
can only
occur in the region $r>a$ of the de Sitter spacetime where all
future-directed
light rays contract. Introducing
$f_A(r) = 1 - r^2 / a^2$ and $f_B(r) = 1 - 2m / r$ in the matching
equation
(\ref{eq2.3}) one gets
\begin{equation} \label{eq2.5}
r_0^3 = 2ma^2.
\end{equation}
Here $a$ is the radius of the de Sitter horizon, $a^2 = 3 / \Lambda
=
3/8 \pi
\rho$, ($\rho$ being the false-vacuum energy density). We assume
that
the value
of $a$ is  a
fixed parameter of our model and that  $a \ll r_0 \ll 2m$.

The two lightlike shells bounding the de Sitter universe behave
identically and
both converge towards $r=0$. Their surface energy densities are
thus
equal
$\sigma_1(r) =  \sigma_2(r)$, and from (\ref{eq2.5}) they are given
by
\begin{equation} \label{eq2.6}
4 \pi r^2 \sigma_1(r) = - m \left(1 - \frac{r^3}{r_0^3}\right).
\end{equation}
This relation shows that the surface energy density is negative
and the value  of the
negative mass of the shells grows to $-m$ as their size $r$ goes to
zero.

In comparison with the FMM--model, where the
transition between the Schwarzschild and de Sitter spacetimes
occurs
instantaneously
along a spacelike shell, we now have a situation which is no longer
homogeneous
as one
moves along  hypersurfaces $r = \mbox{const} < r_0$. This
inhomogeneity can even  be
enhanced  if several bubbles are created and if their boundaries
intersect.

\section{Creation of Multiple de Sitter Bubbles}
\subsection{Conditions for the Intersection of Bubbles}
Consider now the creation of multiple bubbles with
a de Sitter-like phase interior.
If a
couple of
bubbles is created not far from one another  their lightlike
boundaries may
intersect before
reaching the singularity $r=0$. Let us obtain the  conditions when
it
occurs.

The Schwarzschild  metric (\ref{eq1.1}) near $r=0$ can be
approximated by
\[
ds^2 = -\frac{r}{2m} dr^2 + \frac{2m}{r} dt^2 +r^2 d\Omega^2.
\]
Introducing the proper time coordinate $d\tau=-(r/2m)^{1/2}dr$ we
get
\begin{equation} \label{eq3.1}
ds^2 = - d\tau^2 + \left(-\frac{3 \tau}{4 m}\right)^{-\frac{2}{3}}
dt^2 +
	\left(-\frac{3 \tau}{4 m}\right)^{\frac{4}{3}}
	(2m)^2 d\Omega^2.
\end{equation}
The radius $r$ and the proper time $\tau$ are related  as $r^3 = 9
m
\tau^2/2$
and $r$
decreases as $\tau$ increases.

Consider a couple of de Sitter bubbles  created at the vertex
spheres
$M$ and
$N$ of the  same radius
$r_0$ and intersecting at the sphere $P$ of the radius $r_1$ (see
Fig.3).
What happens after the collision of the lightlike shells at the
sphere $P$  is
for the moment left unspecified and will be discussed later.

\begin{figure}
\let\picnaturalsize=N
\def\picsize{12.0cm}
\def\picfilename{fig3.eps}
\ifx\nopictures Y\else{\ifx\epsfloaded Y\else\input epsf \fi
\let\epsfloaded=Y
\centerline{\ifx\picnaturalsize N\epsfxsize \picsize\fi
\epsfbox{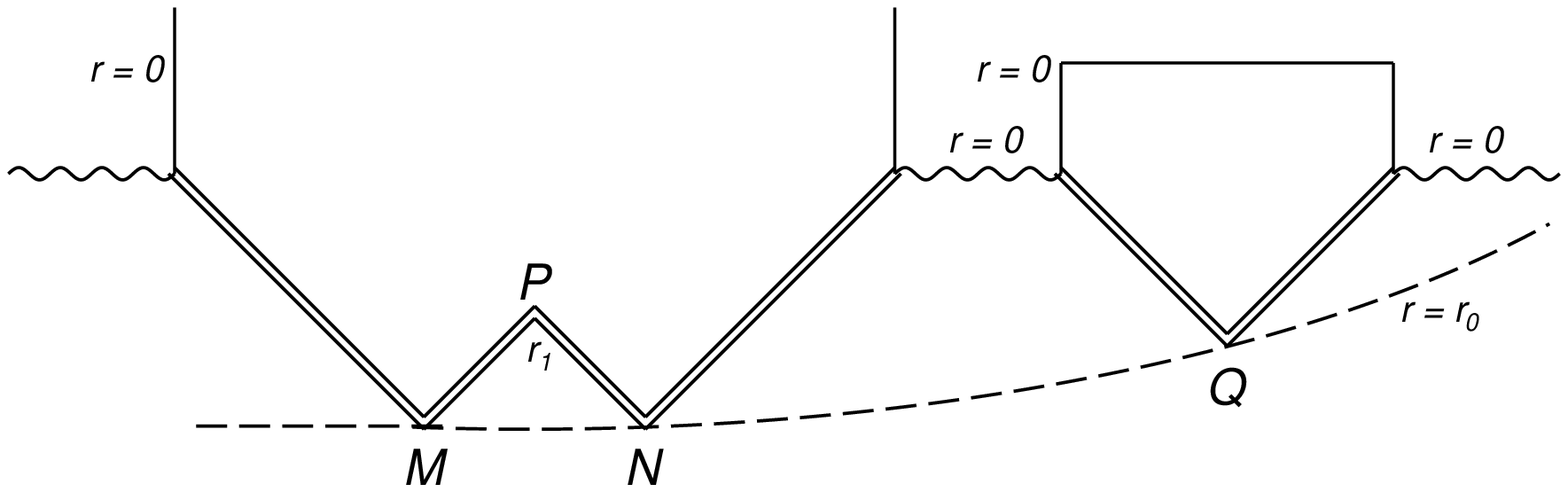}}}\fi
\caption{Multiple bubbles of the de Sitter-like phase creation.}
\end{figure}

Using the metric (\ref{eq3.1}) it is easy to show that  the proper
distance $l$
between two
vertex spheres $M$ and $N$ expressed as the function of   the
intersection
radius   $r_1$ is
\begin{equation} \label{eq3.2}
l = r_0 \left( 1 - \frac{r_1^2}{r_0^2} \right)
	\left( \frac{r_0}{2 m}
\right)^{\frac{1}{2}}=a\left(1-{r_1^2\over
r_0^2}\right),
\end{equation}
while the coordinate $t$ distance is
\begin{equation} \label{eq3.2t}
\Delta t = \frac{r_0^2 - r_1^2}{2m}={a^2\over{r_0^3}}(r_0^2-r_1^2).
\end{equation}
The distance $l$ reaches its maximum value  $l_{\mbox{\scriptsize
max}}$ when
$r_1 = 0$. One has
\begin{equation} \label{3.2f}
l_{\mbox{\scriptsize max}} =
	r_0 \left(\frac{r_0}{2 m}\right)^{\frac{1}{2}} =
	a, \hspace{1cm} \Delta t_{\mbox{\scriptsize max}} =
\frac{r_0^2}{2m}={a^2\over r_0} .
\end{equation}

Two de Sitter universes  created at the vertex spheres $N$ and $Q$,
and separated by
a proper distance $l$ which is
larger than $l_{\mbox{\scriptsize max}}$  remain completely
disconnected. For
$l < l_{\mbox{\scriptsize max}}$ the boundaries of the two bubbles
intersect
before
they reach the singularity. What happens after the intersection
depends upon
the
assumptions made at the collision of the two lightlike shells. In
this Section
we consider the simplest possible scenario, when  the two shells
crossing
one
another have only gravitational  interaction and
remain lightlike after their intersection. The second logically
possible
and more
complicated case  of  merging  shells will be discussed in the next
Section.

\subsection{Crossing of the boundaries}
Consider at first the case  when the in-coming lightlike shells
pass
through
each other and produce two   outgoing lightlike shells. We assume
that after
the collision a new de Sitter universe,
with a different horizon $a'$, is formed. By using the matching
relation
(\ref{eq2.1}) we get
	\begin{equation} \label{eq4.1}
	\left(1 - \frac{2 m}{r_1}\right)
	\left(1 - \frac{r_1^2}{a'^2}\right) =
	\left(1 - \frac{r_1^2}{a^2}\right)^2,
\end{equation}
where $r_1$ is the radius of the intersection sphere. Because the
intersection
takes place inside the gravitational radius of the black hole
($r_1<2m$)  the
collision can only occur if   $r_1$  is larger than $a'$.

Using the dimensionless variables
\begin{equation} \label{eq4.2}
z = \frac{r_0}{a} = \left(\frac{2m}{a}\right)^{\frac{1}{3}},
\hspace{1.5cm} x = \frac{r_1}{a},
\end{equation}
one can rewrite the equation (\ref{eq4.1}) in the following form
\begin{equation} \label{eq4.3}
\left(\frac{a}{a'}\right)^2 = \frac{1}{x^2} + \frac{(1 -
x^2)^2}{x(z^3-x)}.
\end{equation}
According to our assumptions $0\le x\le z$.
For $a=a'$ one has $x=1$. Eq. (\ref{eq4.3}) shows that $x>1$ for
$a<a'$ and
$x<1$ for $a>a'$. In the former case we have  $a'<r_1<a$, while in
the latter
case $a<a'<r_1$ . These two cases correspond to two different ways
of
gluing
the two de Sitter spacetimes shown respectively in Fig.4.a and
Fig.4.b.
A new bubble of false vacuum with a different energy density
appears
between the two initially created false vacuum bubbles. The new
bubble  either
coexists indefinitely with two others [Fig.~4.a] or finally
occupies
the
whole space  [Fig.~4.b]. In the former case  the initial false
vacuum
bubbles
must be nucleated  close enough to one another.
As the limiting value of $r_1$ separating the two cases
is such that $r_1=a=a'$, if follows from (\ref{eq3.2}) that
this occurs whenever the proper distance between the
vertex spheres is smaller than $a(1-a^2/r_{0}^2)$.
In the latter  case, shown in Fig 4.b, the new de Sitter spacetime
is
flatter than the original ones.

\begin{figure}
\let\picnaturalsize=N
\def\picsize{5.0in}
\def\picfilename{fig4.eps}
\ifx\nopictures Y\else{\ifx\epsfloaded Y\else\input epsf \fi
\let\epsfloaded=Y
\centerline{\ifx\picnaturalsize N\epsfxsize \picsize\fi
\epsfbox{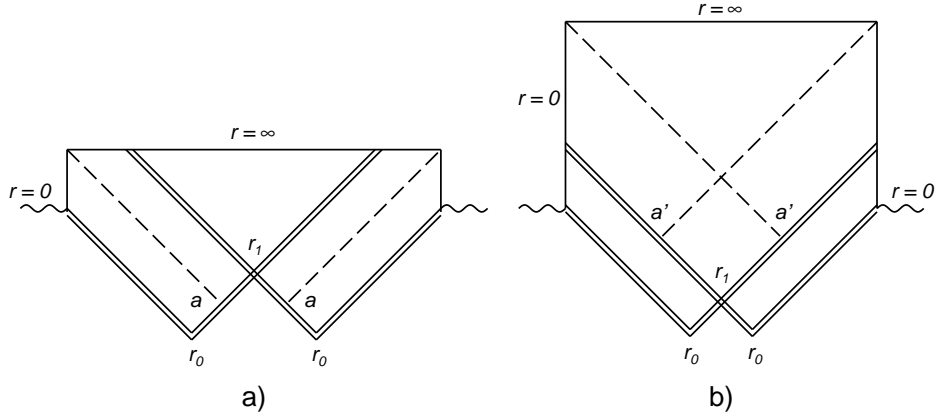}}}\fi
\caption{Free intersection of  lightlike boundaries of two bubbles}
\end{figure}

For symmetry reasons the equations of motion of both shells
as well
as their surface energy density are identical $\sigma'_1(r) =
\sigma'_2(r)$.
The surface energy density of the shells after their intersection
can be
obtained from Eq.(\ref{eq2.4}) and it is of the form
\begin{equation} \label{eq4.4}
4 \pi \sigma'_1(r) = \frac{\zeta r}{2}
	\left(\frac{1}{a^2} - \frac{1}{a'^2} \right),
\end{equation}
This relation shows that the  surface energy density of the shells
after their
intersection is a linear function of the radius  $r$. In the case
where
$a'<r_1<a$,
we have $\zeta = 1$ and the lightlike shells which were initially
contracting
bounce at the collision and expand to infinity [see Fig 4.a]. In
the
second
case [Fig 4.b] we have $\zeta =
-1$ and the lightlike boundaries contract  to zero radius. In both
cases
$\sigma'_1(r)$ is negative as it is
expected to be from the law of conservation of energy at the
collision (the
ingoing
shells have negative energies). For the case shown in Fig.4.b the
mass of the
shells vanishes at the point when the shells cross $r=0$.

\section{Interaction of the Lightlike Boundaries of Two de Sitter
Universes}
\subsection{Merging of the boundaries}
A different situation which may occur at the collision of the two
false vacuum
bubbles is
when their lightlike boundaries interact strongly and merge into a
single
timelike shell (this process
is in some sense analogous to the creation of a massive particle
from
two colliding photons). In that case the two bubbles remain
attached
after
the collision through a spherical surface layer moving with
subluminal
velocity (see Fig.5).

\begin{figure}
\let\picnaturalsize=N
\def\picsize{8.0cm}
\def\picfilename{fig5.eps}
\ifx\nopictures Y\else{\ifx\epsfloaded Y\else\input epsf \fi
\let\epsfloaded=Y
\centerline{\ifx\picnaturalsize N\epsfxsize \picsize\fi
\epsfbox{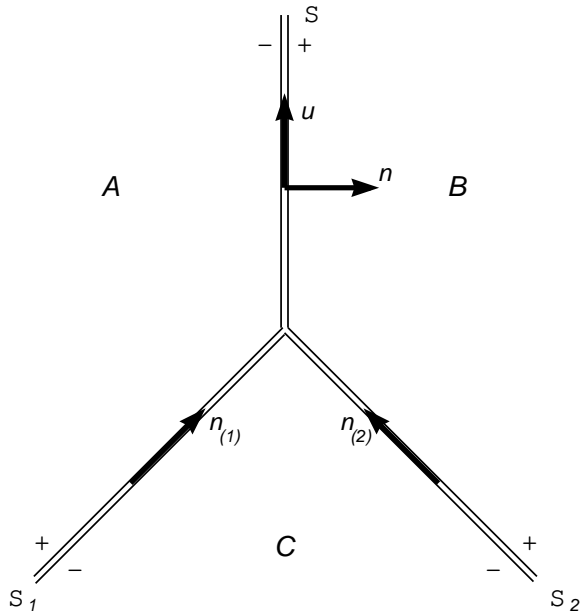}}}\fi
\caption{Merging of two lightlike shells into  one timelike shell.}
\end{figure}

Both in-coming lightlike shells are  contracting, so that   the
radius of
collision $r_1$ is smaller than $r_0$. The metrics in the regions
$A$
and $B$
shown in Fig.5 are  de Sitter metrics, while the metric in the
region
$C$ is
the Schwarzschild one. The corresponding metric functions $f(r)$
are
\begin{equation} \label{eq4.5}
f_A(r) = f_B(r) = 1 - \frac{r^2}{a^2},
\hspace{1.5cm} f_C(r) = 1 - \frac{2m}{r}.
\end{equation}

Let us obtain the expression for the parameters of the timelike
shell
in terms
of the parameters for colliding null shells. For this purpose we
use
the
matching condition (\ref{eqA.13})
\begin{equation} \label{eq4.5a}
	\left(\dot{r_1} + \varepsilon_A
\sqrt{f_A(r_1)+\dot{r_1}^2}\right)
	\left(\dot{r_1} - \varepsilon_B
\sqrt{f_B(r_1)+\dot{r_1}^2}\right) =
	-f_C(r_1).
\end{equation}
(This relation as well as other useful formulas for moving and
colliding shells
are collected in the Appendix.) Here $r_1$ is the radius of the
collision
sphere, and the parameter $\varepsilon$ which enters this
expression
is defined
as $\varepsilon =
\mbox{sign}(n^\alpha \partial_\alpha r)$.
Since the spacetime domains bordering the timelike
shell are identical, we must take  $\varepsilon_A = -
\varepsilon_B$,
otherwise there is no shell
(see the Eq.(A.5) of the Appendix).
Eq.(\ref{eq4.5a})
implies
\begin{equation} \label{eq4.9}
\dot{r_1}^2 = - \frac{\left(f_A(r_1) + f_C(r_1)\right)^2}{4
f_C(r_1)}.
\end{equation}
On the other hand for the timelike shell one has (see Eq.(A.5))
\begin{equation} \label{eq4.6}
\frac{M(r)}{r} = 2 \varepsilon_A
	\left(f_A(r) + \dot{r}^2 \right)^{\frac{1}{2}},
\end{equation}
where $M(r) = 4 \pi r^2 \sigma(r)$ is the inertial mass of the
timelike shell
and $\sigma(r)$
its energy  surface density.
By combining the relations (\ref{eq4.9}) and (\ref{eq4.6}) we get
the initial mass $M(r_1)$ of the timelike shell at the moment of
the
collision
\begin{equation} \label{eq4.10}
M(r_1) = \frac{2 M_1(r_1)}{\left|f_C(r_1)\right|^{\frac{1}{2}}}.
\end{equation}
Here $M_1(r_1) = 4 \pi r_1^2 \sigma_1(r_1)$ is the mass of the
lightlike shells
at the
collision. As expected from the conservation of energy, it follows
from
(\ref{eq4.10}) that  the masses of the ingoing lightlike shells and
of the
outgoing timelike
shell have the same negative sign (recall that the lightlike shells
have
negative energy densities).

The further evolution of the timelike shell is given by  equation
(\ref{eq4.6}). It can be rewritten in the form
\begin{equation} \label{eq4.7}
\dot{r}^2 + V(r) = -1,
\end{equation}
where $V(r)$ is an effective potential given by
\begin{equation} \label{eq4.8}
V(r) = - \frac{r^2}{a^2} - 4 \pi^2 r^2 \sigma^2(r).
\end{equation}
Eq. (\ref{eq4.7}) shows that the motion is only possible when
$V(r) \le
-1$.

To study this equation it is convenient to use  the dimensionless
radius
$y=r/a$ and time $T=\tau/a$ as well as the variables $x$ and $z$
already
defined by
(\ref{eq4.2}). ( $x$ is an initial value of $y$ and $dy/dT =
\dot{r}$.) In
these variables the equation of   motion (\ref{eq4.7}) takes the
form
\begin{equation} \label{eq4.12}
\left(\frac{dy}{dT}\right)^2 + V(y,x,z) = -1,
\end{equation}
where $z$ is a given quantity and $x$ a free parameter such that $z
\gg 1$ and
$x \in
[0,z]$. The initial mass (\ref{eq4.10}) of the timelike shell can
be
rewritten
as
\begin{equation} \label{eq4.13}
M(r_1) = a F(x, z)
\end{equation}
with
\begin{equation} \label{eq4.14}
F^2(x,z) = \frac{(z^3 - x^3)^2 x}{z^3 - x}.
\end{equation}

\subsection{Evolution of a Merging Boundary}
The motion of the timelike shell depends on the equation of state
for
the
matter forming the shell.  For
simplicity we assume a dust-like equation of state.  Under this
assumption
$\sigma r^2 = \mbox{const} < 0$ along the
shell and the mass of the shell is conserved and coincides with
$M(r_1)$ given by Eqs. (\ref{eq4.13}) --- (\ref{eq4.14}).

For a  dust-like equation of state the potential $V$ is
\[
V(r) = - \frac{r^2}{a^2} - \frac{M^2(r_1)}{4 y^2},
\]
and  written in terms of the dimensionless variables it takes
the form
\begin{equation} \label{eq4.15}
V(y,x,z) = - y^2 - \frac{F^2(x,z)}{4 y^2}.
\end{equation}
If the initial condition is chosen so that   $T = 0$
when $y = x$, the solution of Eq.(\ref{eq4.12}) is given in the
following
implicit form
\begin{equation} \label{eq4.16}
T = \frac{1}{2} \log
	\left|\frac{2y^2-1+(4y^4-4y^2+F^2)^{\frac{1}{2}}}{
	2x^2-1+(4x^4-4x^2+F^2)^{\frac{1}{2}}}\right|.
\end{equation}
The solution depends on $x$ (the initial value of $y$) and on the
parameter
$z$. We consider a black hole of large mass, so that we have
$z\gg1$.

Motion of the timelike shell is possible only in the region where
$V(y,x,z) \le -1$. Denote by  $y_m$ the value of $y$ where the
potential $V$
reaches its maximum value $V_m = V(y_m)$.
Eq. (\ref{eq4.15}) implies that
\begin{equation} \label{eq4.17}
y_m = \frac{F^{\frac{1}{2}}(x,z)}{\sqrt{2}}, \hspace{1.5cm} V_m = -
F(x,z).
\end{equation}
It is easy to show that there exist two values of $x$ ($x_1$ and
$x_2$) for
which $V_m = -1$. For $z\gg 1$ one has
\begin{equation} \label{eq4.18a}
x_1 = \frac{1}{z^3} + o\left(\frac{1}{z^7}\right),\hspace{1cm} x_2
=
z -
\frac{1}{3 z} + o\left(\frac{1}{z^5}\right) .
\end{equation}
$V_m < -1$ for $x \in [x_1, x_2]$ (the curve $B$ in the Fig.6) and
$-1 < V_m <
0$ for $x \in [0,x_1]$ or $x \in  [x_2, z]$ (the curves $A$ and $C$
in the
Fig.6).

\begin{figure}
\let\picnaturalsize=N
\def\picsize{8.0cm}
\def\picfilename{fig6.eps}
\ifx\nopictures Y\else{\ifx\epsfloaded Y\else\input epsf \fi
\let\epsfloaded=Y
\centerline{\ifx\picnaturalsize N\epsfxsize \picsize\fi
\epsfbox{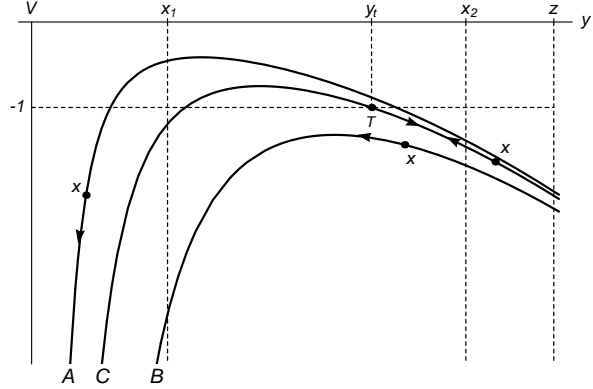}}}\fi
\caption{The potential $V(y,x,z)$ for different values of the
parameter $x$.}
\end{figure}

For a fixed parameter $z$ the evolution of the shell depends on the
initial
condition $x$ (the initial radius of the timelike shell).
A timelike shell   is created
at a stage of contraction, so that $(dy/dT)_{T=0}=\dot{r_1}<0$.
Under this
condition only two qualitatively different types of motion are
possible:

(1)  $0 < x \le x_2$. In this case the timelike shell monotonically
contracts
from its initial
radius $r_1$ down to zero\footnote{This is evident for $x_1<x<x_2$
because there
are no turning points. For $0<x<x_1$  there are turning points, but
the
contraction begins at the initial value  $y=x$ lying to the left of
the turning
points.}. The two de Sitter universes which were initially attached
through the timelike shell become disconnected after the time
interval
$T_{\mbox{\scriptsize sep}}$
\begin{equation} \label{eq4.19}
T_{\mbox{\scriptsize sep}} = \frac{1}{2} \log
	\left| \frac{F-1}{2x^2-1+(4x^4-4x^2+F^2)^{\frac{1}{2}}}
\right|.
\end{equation}
For $x \in [0,x_1]$ (the curve $A$ in Fig.6) the separation time is
less
than $z^{-6}a$ and the bubbles separate almost immediately
after their intersection. For $x \in [x_1,x_2]$, curve $B$ of
Fig.6,
the time
of the separation
can reach values of the order of  $a$.

(2) $x_2 < x < z$.  The trajectory admits a turning point $V(y_t) =
-1$ at the
value $y=y_t$ \begin{equation}
y_t^2 = \frac{1}{2} \left[1+\left(1-F^2\right)^{\frac{1}{2}}\right]
{}.
\end{equation}
In that case the timelike shell first contracts, bounces at $y_t$,
and then
expands to infinity (the curve
$C$ in Fig.6). The two de Sitter universes remain connected through
a
spherical
shell
which has a radius that bounces and increases to infinity. The time
which is
needed for the
shell to bounce from its initial radius can be estimated as $T \sim
\log z$.

The conformal Penrose-Carter diagrams for these two qualitatively
different
cases are shown in Fig.7. The Fig.7.a  illustrates the evolution of
a
timelike
shell with the initial condition  $y=x \in [0,x_2]$ at the moment
of
its
formation. The contraction of the timelike shell results in the
separation of
two de Sitter-like universes at the moment $P$.    Fig.7.b
illustrates the
evolution of a timelike shell with the initial condition  $y=x \in
[x_2, z]$.
The timelike shell changes its contraction into expansion.  The
two de
Sitter-like  universes remain
connected.

The condition  $x_2 < x < z$ which guarantees that two
bubbles form the
same de~Sitter-like universe after merging of their boundaries
can be rewritten in terms of the
separation
between the vertex spheres of the bubble nucleation. First we
remark
that this
condition written in  dimensional form is $r_0-a^2/3r_0<r_1<r_0$.
(We use
here the relation (\ref{eq4.18a}) valid for $z\gg 1$). Using the
relations
(\ref{eq3.2}) and (\ref{eq3.2t}) we get that this condition is
equivalent to
\begin{equation}
0<l<{a\over 3}\left(a\over{r_0}\right)^2,\hspace{1cm}0<\Delta
t<{a\over
3}\left(a\over{r_0}\right)^3.
\end{equation}
Since $a/r_0\ll 1$ two bubbles form a unique de Sitter universe
only if their
vertices are extremely close one to another. In other words, in the
framework of
the chosen model the creation of multiple disconnected de
Sitter-like
universes
is the most plausible process.

\begin{figure}
\let\picnaturalsize=N
\def\picsize{13 cm}
\def\picfilename{fig7.eps}
\ifx\nopictures Y\else{\ifx\epsfloaded Y\else\input epsf \fi
\let\epsfloaded=Y
\centerline{\ifx\picnaturalsize N\epsfxsize \picsize\fi
\epsfbox{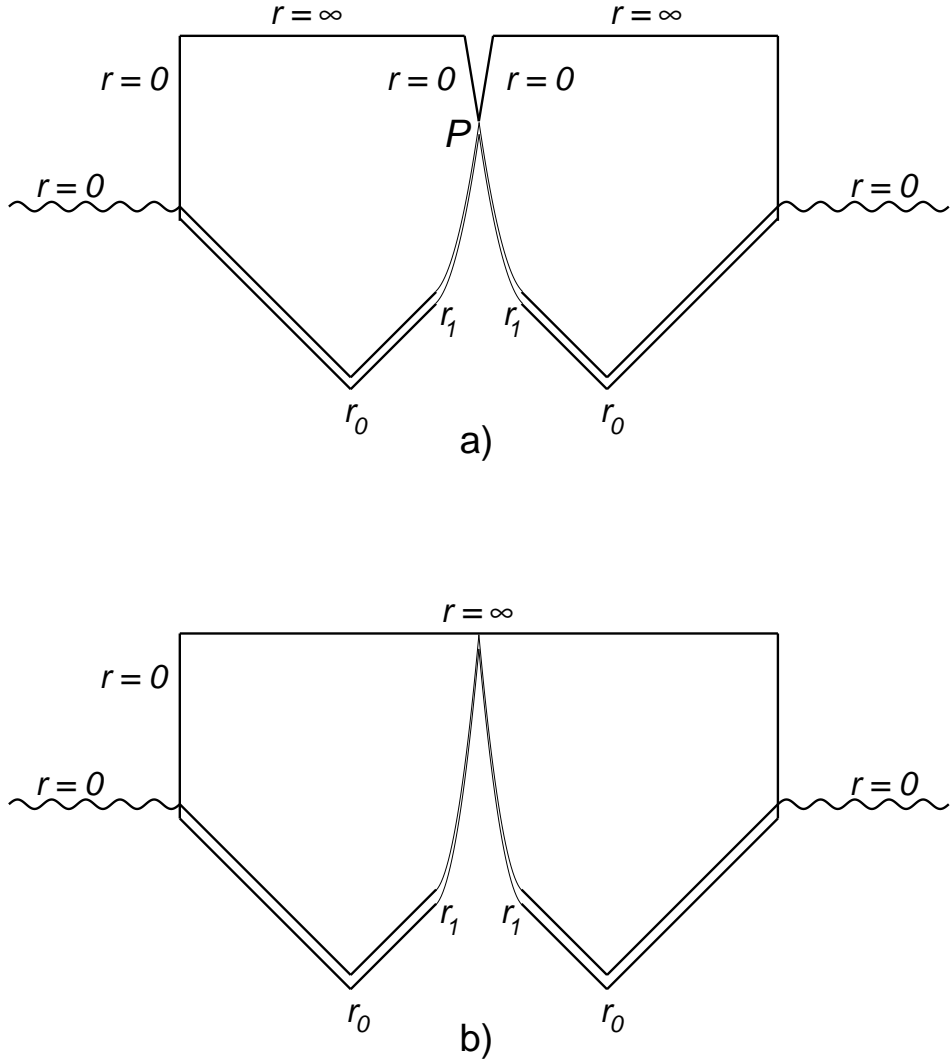}}}\fi
\caption{Conformal Penrose-Carter diagrams for a spacetime with two
de
Sitter-like phase bubbles with merging boundaries for different
distances
between bubbles at the moment of creation. In both figures two
double
lines
beginning at the radius $r_1$ and lying to the future represent the
same
timelike shell and must be identified.}
\end{figure}

\section{Concluding Remarks}
We discuss now some physical consequences of the possible multiple
de
Sitter-like universe formation inside a black hole. The above
consideration
indicates that the nucleation of bubbles of the de Sitter-like
phase
might
result in the formation of disconnected worlds as soon as $\Delta
t{\
\lower-1.2pt\vbox{\hbox{\rlap{$>$}\lower5pt\vbox{\hbox{$\sim$}}}}\
}a^4/r_0^3$.
It means that during the time  $t_{evap}\sim t_{Pl}(m/\mu_{Pl})^3$
of
the
quantum evaporation of a  black hole of mass $m$ there might be
formed as many
as
\begin{equation}\label{eq5.1}
N_{max}\sim \left( l_{Pl}\over{a}\right) ^2 \left( m\over
{\mu_{Pl}}\right)^4
\end{equation}
new de Sitter-like disconnected universes. ($t_{Pl}$, $l_{Pl}$, and
$\mu_{Pl}$
are Planckian time, length, and mass, respectively.) For  $a\sim
l_{Pl}$ when
the curvature of a newly created de Sitter world is Planckian,
$N_{max}\sim
\left( m/ {\mu_{Pl}}\right)^4$. Remarkably this quantity is much
larger
than the dimensionless Bekenstein-Hawking entropy of a black hole
$S^{BH}$ or
the number of emitted quanta of the Hawking radiation
$N\sim \left( m/ {\mu_{Pl}}\right)^2$ (both are of
the same
order of magnitude).

The number of different possibilities for the distribution of $N$
identical
particles in $N_{max}$ 'boxes' (representing newly created worlds)
is
$N_{max}!/(N_{max}-N)! N!$. If these possibilities are equally
likely
then the corresponding entropy $S$ is
\begin{equation}\label{eq5.2}
S=\log [N_{max}!/(N_{max}-N)! N!] .
\end{equation}
Using the relation $\log N! \approx N\log (N/e)$ we get for $N\ll
N_{max}$
\begin{equation}\label{eq5.3}
S\approx N\log (N_{max}/N ).
\end{equation}
In other words for $N_{max}\gg N$ the entropy connected with all
possible
distributions of the created Hawking quanta over the newly born
worlds
inside a
black hole is greater than the entropy of these quanta. It means
that
entropy considerations in principle do not exclude the situation
where
the internal state of each newly created universe is pure while the
ensemble of
universes is described by a density matrix. This remark might
have quite
interesting applications to the information loss puzzle.

To summarize, we proposed a simple model for  the black hole
interior. In the
framework of this model we showed that at least kinematically the
creation of
many separated universes is the most probable process. Certainly
the
model
itself contains a lot of simplifications. But the obtained results
indicate
that multiple universe formation in the black hole interior should
be
taken
seriously in the discussion and classification of different
non-singular black
hole models.

\vspace{.5cm}
{\bf Acknowledgments}:\ \ The authors thank to Werner Israel,
Slava Mukh\-anov, Eric
Poisson, and
Roberto Balbinot for fruitful discussions. The authors are grateful
to Andrei
Frolov for his help in the preparation of the paper. This work was
partly
supported  by the Natural Sciences and Engineering Research Council
of Canada.

\appendix
\section{Timelike and Lightlike Shells in a Spherically-Symmetric
Spacetime}
\setcounter{equation}{0}
\renewcommand{\theequation}{A.\arabic{equation}}

In this appendix we derive the matching relations which has to be
fulfilled
when two
ingoing lightlike shells merge into a single timelike shell. We
only
consider
the case of
concentric spherical shells that move radially. The three spacetime
domains
bounded by
the shells are static and spherically symmetric with line elements
of
the
following form
\begin{equation} \label{eqA.1}
ds^2 = -f(r) dt^2 + f^{-1}(r) dr^2 + r^2(d\theta^2 + \sin^2\theta
d\phi^2).
\end{equation}
We call $A$, $B$, $C$ the three spacetimes, $\Sigma_1$, $\Sigma_2$
the
lightlike shells
and $\Sigma$ the timelike shell [See Fig.5].

Before deriving the matching relation at the intersection let us
recall the
main equations
describing timelike and lightlike shells (for more details see
\cite{BaIs:91}).

We call $u = d/d\tau$ the normalized 4-velocity and $n$ the unit
normal along
the timelike
surface $\Sigma$. With the line element (\ref{eqA.1}) the
components
of these
vectors are
\begin{equation} \label{eqA.2}
u^\alpha = \left( \varepsilon_1 f^{-1}(r) \sqrt{f(r) + \dot{r}^2},
		\dot{r}, 0, 0 \right),
\end{equation}
\begin{equation} \label{eqA.3}
n^\alpha = \left( \varepsilon_2 \dot{r} f(r),
		\varepsilon \sqrt{f(r) + \dot{r}^2}, 0, 0 \right),
\end{equation}
with $\varepsilon_1, \varepsilon_2 = \pm 1$, $\varepsilon =
\varepsilon_1
\varepsilon_2 =
\mbox{sign}(n^\alpha \partial_\alpha r)$ and $\dot{r} = d/d\tau$.
The
induced
metric of
$\Sigma$ is equal to
\begin{equation} \label{eqA.4}
ds^2 = - d\tau^2 + r^2(\tau) (d\theta^2 + \sin^2\theta d\phi^2).
\end{equation}
A spherically symmetric timelike shell is a 2-dimensional perfect
fluid with
surface energy
$\sigma$ and surface pressure $p$. Let us call ${\cal A} = 4 \pi
r^2$
the area
and $M =
\sigma {\cal A}$ the internal mass. Spherical symmetry reduces the
number of
independent
equations describing the motion of the shell to the following
two :
\begin{equation} \label{eqA.5}
- \frac{M}{r} = \left[ \varepsilon \sqrt{f(r) + \dot{r}^2} \right],
\end{equation}
\begin{equation} \label{eqA.6}
\frac{dM}{d\tau} = - p \frac{d{\cal A}}{d\tau} +
	\left[ T_{\alpha \beta} u^\alpha n^\beta \right] .
\end{equation}
Here $T_{\alpha \beta}$ is the stress-energy tensor of the external
medium and the
square bracket $[ \hspace{3pt} ]$ represents the jump of the
enclosed
quantity across the
surface, i.e. $[F] = F_+ - F_-$.  It is assumed that the normal $n$
is
directed toward the $+$   side.

Eq.(\ref{eqA.5}) is the equation of motion of the shell and
Eq.(\ref{eqA.6}) the equation of
conservation of energy. To specify the problem one needs
also to give an equation of state.
If, for instance, one
takes $p = (\alpha-1)\sigma$ with $\alpha = \mbox{const}$ and if
$[
T_{\alpha\beta}
u^\alpha n^\beta ] = 0$ one gets from (\ref{eqA.6})
\begin{equation} \label{eqA.7}
\sigma r^{2\alpha} = \mbox{const}.
\end{equation}
This relation allows one to define $M=M(r)$ which
enters Eq. (\ref{eqA.5}).

The description of a null shell is very different from a timelike
one and in some sense
simpler because the equation of motion is fixed.
What makes a null hypersurface  so peculiar is that
its normal vector is at the same time tangent to it and that its
induced metric is degenerate.

Let us call $\Sigma_i$ with $i = 1,2$ the two ingoing null shells.
The basis
vectors $e_{(A)}
= \partial / \partial \xi^A$ where $\xi^A = (\theta,\phi)$ are
tangent to the
shells and one
takes $n_{(i)} = \zeta_i \partial / \partial r$ as the null vector
tangent to
the null generators
of $\Sigma_i$. Here $\zeta_i = +1 (-1)$ whenever $r$ increases
(decreases)
toward the
future along the null generators. As $n_{(i)}$ is tangent to
$\Sigma_i$, the
``extrinsic''
curvature defined by
\begin{equation} \label{eqA.8}
{K_i}_{AB} = - n_{(i)} \cdot \frac{\delta e_{(A)}}{\delta \xi^B},
\end{equation}
where $\delta / \delta \xi^B$ is the 4-dimensional covariant
derivative, an
intrinsic
property of the shell which actually describes the behavior of its
null
generators. For
instance, the trace $K_i = g^{AB} {K_i}_{AB}$ represents their
expansion rate
and is equal
to
\begin{equation} \label{eqA.9}
K_i = \frac{2}{r} n_{(i)}^\alpha \partial_\alpha r.
\end{equation}
Finally when no energy is transferred to the shell, i.e. $[
T_{\alpha\beta}
n_{(i)}^\alpha
n_{(i)}^\beta ] = 0$, the surface stress-energy tensor of a
spherical
null
shell is
characterized only by a  surface energy density
 $\sigma_i(r)$, which is
given by
\begin{equation} \label{eqA.10a}
4 \pi r^2 \sigma_1(r) = \frac{\zeta_1 r}{2} \left[f_A(r) -
f_C(r)\right],
\end{equation}
\begin{equation} \label{eqA.10b}
4 \pi r^2 \sigma_2(r) = \frac{\zeta_2 r}{2} \left[f_B(r) -
f_C(r)\right].
\end{equation}

Let us now find the matching relations at the intersection of the
shells. This
intersection is
in fact a 2-sphere $S$ with radius $r_1$ and at any point of $S$ we
can write
the following
decompositions
\begin{equation} \label{eqA.11a}
n_{(1)}^\alpha = (u^\alpha + n^\alpha)/\eta_1 \sqrt{2},
\end{equation}
\begin{equation} \label{eqA.11b}
n_{(2)}^\alpha = (u^\alpha - n^\alpha)/\eta_2 \sqrt{2},
\end{equation}
where $\eta_1$ and $\eta_2$ are arbitrary positive scalars.
Furthermore the
4-vectors
$(n_{(i)}, e_{(A)})$ form a basis at any point of $S$ and we have
the
completeness relation
\begin{equation} \label{eqA.12}
g^{\alpha \beta} = g^{AB} e_{(A)}^\alpha e_{(B)}^\beta -
	\frac{2 n_{(1)}^{(\alpha} n_{(2)}^{\beta)}}{n_{(1)} \cdot
n_{(2)}}.
\end{equation}
In order to get the matching relation we use the fact that $K_i$
are
intrinsic
quantities and
express the product $K_1 K_2$ in two different manners using the
spacetime
domains
$A$, $B$ and $C$. First using (\ref{eqA.2}), (\ref{eqA.3}),
(\ref{eqA.9}) and
(\ref{eqA.11a})
one gets
\[
K_1 K_2 = \frac{2}{r_1^2 \eta_1 \eta_2}
	\left(\dot{r_1} + \varepsilon_A
\sqrt{f_A(r_1)+\dot{r_1}^2}\right)
	\left(\dot{r_1} - \varepsilon_B
\sqrt{f_B(r_1)+\dot{r_1}^2}\right).
\]
Second using (\ref{eqA.9}) and (\ref{eqA.12}) in sector $C$ one
gets
\[
K_1 K_2 = - \frac{2 n_{(1)} \cdot n_{(2)}}{r_1^2} f_C(r_1).
\]
One then immediately derives the matching relation
\begin{equation} \label{eqA.13}
	\left(\dot{r_1} + \varepsilon_A
\sqrt{f_A(r_1)+\dot{r_1}^2}\right)
	\left(\dot{r_1} - \varepsilon_B
\sqrt{f_B(r_1)+\dot{r_1}^2}\right) =
	-f_C(r_1).
\end{equation}
This equation gives the initial velocity $\dot{r_1}$ of the
timelike shell after the
merging of the lightlike shells. When this result is inserted in
(\ref{eqA.5}) one gets the initial mass of the timelike shell.


\begin {thebibliography}{99}
\bibitem{ftn1} In this paper we consider a generalization of the
model proposed
in Refs.\cite{FrMaMu:89,FrMaMu:90}. For briefness we refer to this
model as to
FMM-model.
\end{thebibliography}
\end{document}